# Magnetic MoS$_2$ Interface Monolayer on CdS Nanowire by Cation Exchange

Chih-Shan Tan[1], Yu-Jung Lu[2], Chun-Chi Chen[3], Pei-Hsuan Liu[1], Shangjr Gwo[2], Guang-Yu Guo[4,5,*], Lih-Juann Chen[1,*]

[1]Department of Materials Science and Engineering, National Tsing Hua University, Hsinchu, Taiwan, 30043, R.O.C.

[2]Department of Physics, National Tsing Hua University, Hsinchu, Taiwan, 30043, R.O.C.

[3]National Nano Device Laboratories, National Applied Research Laboratories, 26, Prosperity Road I, Hsinchu, Taiwan, 30078, R.O.C.

[4]Department of Physics, National Taiwan University, Taipei, Taiwan, 10617, R.O.C

[5]Physics Division, National Center for Theoretical Sciences, Hsinchu, Taiwan, 30078, R.O.C

**ABSTRACT:** MoS$_2$ atomic layers have recently attracted much interest because of their two-dimensional structure as well as tunable optical, electrical, and mechanical properties for next generation electronic and electro-optical devices. Here we have achieved facile fabrication of MoS$_2$ thin films on CdS nanowires by cation exchange in solution at room temperature and importantly observed their extraordinary magnetic properties. We establish the atomic structure of the MoS$_2$/CdS heterostructure by taking atomic images of the MoS$_2$/CdS interface as well as performing first-principles density functional geometry optimizations and STEM-ADF image simulations. Furthermore, our first principles density functional calculations for the MoS$_2$/CdS heterostructure reveal that the magnetism in the MoS$_2$/CdS heterostructure stems from the ferromagnetic MoS$_2$ monolayer next to the MoS$_2$/CdS interface. The ferromagnetism is attributed to the partial occupation of the Mo $d_{x^2-y^2}/d_{xy}$ conduction band in the interfacial MoS$_2$ monolayer caused by the mixed covalent-ionic bonding among the MoS$_2$ and CdS monolayers near the MoS$_2$/CdS interface. The present findings of the ferromagnetic MoS$_2$ monolayer with large spin polarization at the MoS$_2$-semiconductor interface suggest a new route for fabrication of the transition metal dichalcogenide-based magnetic semiconductor multilayers for applications in spintronic devices.

# INTRODUCTION

Transition metal dichalcogenide (TMD) monolayers [1-4], especially that of molybdenum disulfide (MoS$_2$) [5-8], have recently attracted much interest because of their controllable optical, electrical, and mechanical properties for next generation devices. Bulk MoS$_2$ consists of sandwich-like S-Mo-S monolayers in which each Mo atom is connected to six S atoms with covalent bonds. The sandwich MoS$_2$ monolayers are bound together by weak van der Waals forces. The band structure of MoS$_2$ undergoes a remarkable change from indirect band gap of ~1.2 eV in the bulk to direct one of ~1.8 eV for single monolayer.[9] MoS$_2$ nanowires exhibit excellent electrochemical performance in lithium ion batteries.[10] Recent researches have been focused on the excellent optical and electrical properties of MoS$_2$ and some are also related to the novel magnetic properties of MoS$_2$. For example, it has been reported that the valley magnetic moment of the MoS$_2$ monolayer could be electrically tuned [11] and that zigzag and armchair nano-ribbons of MoS$_2$ could have metallic (ferromagnetic) and semiconducting (nonmagnetic) behavior [12-15]. It has also been predicted that large spin-polarization could be induced in the interfacial MoS$_2$ monolayer in the ferromagnetic Fe$_4$N/MoS$_2$ bilayer due to the magnetic proximity effect.[16] Spintronic devices are being developed with vigorous efforts [17-18] and spin transistor [5, 19-20] is believed to be the most promising next generation spintronic device.

Although a number of researches were directed to form MoS2 films with atomic thickness, scarce attention has been paid to the formation process of MoS$_2$ films and the bonding and physical properties of the MoS$_2$ layer with other semiconductor materials. Here we use the cation exchange to grow epitaxial MoS$_2$ layers on the surface of CdS nanowires (NWs) at room temperature. Remarkably, we discover that the MoS$_2$/CdS heterostructures are ferromagnetic. Our first-principles density functional calculations reveal that the magnetism originates from the MoS$_2$ monolayer at the MoS$_2$/CdS interface with a spin magnetic moment of ~0.5 $\mu_B$ on the Mo atom and a large spin-polarization (~63 %) of the electronic states at the Fermi level. The occurrence of the magnetization in the interfacial MoS$_2$ monolayer is attributed to the Mo cation sharing of its d electron toward Cd cation. As MoS$_2$ is considered to be an excellent material for transistors, [5] the present findings of the itinerant magnetism in the MoS$_2$ monolayer at the MoS$_2$–semiconductor interface may open up a new opportunity for the fabrication of the magnetic two-dimensional (2-D) TMD materials on three-dimensional (3-D) nanostructures for application in spin transistors and other semiconductor spintronic devices.

**EXPERIMENTAL AND COMPUTATIONAL METHODS**

**CdS NW Synthesis:** CdS NWs were grown by using the method reported in ref. 21.

**Cation Exchange Transformation:** CdS NWs were dipped into ethylene glycol (99.5%, Sigma-Aldrich) containing 0.1 M molybdenum (V) chloride (Alfa Aesar, 99.6%) at room temperature. After 6 hr, the CdS NWs completely transformed to MoS$_2$ NWs. We can change the MoS$_2$ shell thickness by varying the reaction time.

**Instrumentation:** FE-SEM images were obtained using a FEI Helios 1200$^+$ FE-SEM. TEM images were obtained using JEOL ARM200F and 2010F electron microscopes. Raman measurements were carried out with Horiba Jobin Yvon, LABRAM HR 800 UV. The magnetic measurements were conducted using a SQUID (MPMS XL-7).

**STEM Image Simulations:** STEM-ADF image simulations of the model $MoS_2$/CdS heterostructures were performed with software QSTEM. [22] The input parameters were set according to our experimental conditions including probe size, convergence angle and acceptance angle of the ADF detector. For enhancing the contrast of the sulphur atoms, the medium-range ADF mode was selected instead of high-angle ADF mode by proper adjustment of the camera length. The detector angle is at a pivot condition (with 20 to 60 mard) between the medium-range ADF mode and the high-angle ADF mode.

**Density Functional Calculations:** Several initial atomic models of the $MoS_2$/CdS heterostructure were constructed by different combinations of the $MoS_2$ ($00\bar{2}$) and CdS (002) thin layers. First-principles geometry optimizations for these initial atomic models of the $MoS_2$/CdS heterostructure were performed based on density function theory (DFT) with the generalized gradient approximation (GGA) of the Perdew–Burke–Ernzerhof form. [23] The optimized $MoS_2$/CdS heterostructure models were then used to simulate the STEM-ADF image. The atomic model that produces the image best fitting to the real STEM-ADF image is displayed in Figure 1, and also used in the subsequent GGA calculations of the electronic and magnetic properties of the $MoS_2$/CdS heterostructure. First-principles DFT-GGA calculations were carried out by using the plane wave norm conserving pseudopotentials method, as implemented in the Cambridge Serial Total Energy Package (CASTEP). The plane wave basis-set cut-off energy is 720 eV. We used a fine k-point mesh of 5×8×1 for the Brillouin zone throughout.

**RESULTS AND DISCUSSION**

Cation exchange is an ionic reaction process for III-V and II-VI semiconductors and metal-organic frameworks to change the components and structures [24-26] as well as a convenient method for fabricating optical and electrical devices. [27] In a previous work, ion exchange was used to convert CdS nanowire templates to $Cu_2S$-$Ag_2S$ superlattice p-n heterojunction NWs due to ionic bonding preferences between different cation and anion.[21] In the present investigation, $MoS_2$ layers were grown on CdS nanowires by cation exchange and the magnetic properties of $MoS_2$ (shell)/CdS (core) NWs were explored. CdS NWs were dipped in a molybdenum ionic solution at room temperature (RT) and atomic layers of $MoS_2$ were grown on CdS NWs. By control of the reaction time, different thicknesses, from one to tens of atomic layers, of $MoS_2$ can be grown on CdS NW as a shell by cation exchange. Finally, the CdS NWs were completely transformed to $MoS_2$ NWs after several hours.

The ex situ Raman spectra of cation exchange from CdS NWs to $MoS_2$ NWs with different reaction times up to 6 hr were monitored. By identifying the $E_{2g}^1$ (in plane vibration) and $A_{1g}$ (out of plane) peaks, the gradual formation of $MoS_2$ structure with increasing reaction time can be inferred (Supplementary Material Figure S1 and reference 28). The data exhibit clearly the 302 $cm^{-1}$ peak, which is the first order longitudinal optical mode (1LO) of CdS. After cation exchange, the CdS

1LO peak shifts gradually with strain and eventually disappears. On the other hand, $E_{2g}^1$ and $A_{1g}$ signals of MoS$_2$ increase gradually. It is difficult to identify the actual location of the $E_{2g}^1$ peak at 376 cm$^{-1}$ near the CdS peaks until the completion of the reaction. The $A_{1g}$ peaks during the cation exchange from CdS to MoS$_2$ NWs are 408 cm$^{-1}$ (2 min), 405 cm$^{-1}$ (3 min), 407 cm$^{-1}$ (4 min), 407 cm$^{-1}$ (5 min), 405 cm$^{-1}$ (10 min), 407 cm$^{-1}$ (30 min), 407 cm$^{-1}$ (1 hr), 406 cm$^{-1}$ (2 hr), and 406 cm$^{-1}$ (6 hr). It indicates that the stress and strain are varied with reaction time and MoS$_2$ thickness. SEM images are shown in supplementary material Fig. S2.

Bright field transmission electron microscopy (BF-TEM) and scanning transmission electron microscopy-energy dispersive spectroscopy (STEM-EDS) were carried out to obtain structural information on the MoS$_2$/CdS interface. We analyzed the MoS$_2$ on CdS NW surface after 3-10 min of cation exchange reaction by BF-TEM (Figure 1a). STEM imaging and EDS mapping of Mo, Cd, and S (Figure 1g), the core-shell structure of CdS-MoS$_2$ is evident. Scanning transmission electron microscopy annular dark-field (STEM-ADF) image (Figure 1b) further reveals that MoS$_2$ (00$\bar{2}$) is connected with CdS (002) (Figure 1b). The STEM-ADF image matches rather well with the simulated image (Figure 1c) using the atomic model described below (Figure 1d). We built many atomic models for the MoS$_2$/CdS heterostructure and then performed first-principles geometry optimization calculations for these models which were used in the STEM-ADF image simulations by QSTEM.[22] The MoS$_2$/CdS heterostructure shown in Figure 1d is the heterostructure model with simulated STEM-ADF image (Figure 1c) fitted best to the real STEM ADF image (Figure 1b).

From the STEM-ADF (Figure 1e) and ABF (Figure 1f) images, with magnified images as insets, the hexagonal structure of $MoS_2$ along [001] direction is evident. Both CdS and $MoS_2$ are of hexagonal close-packed (hcp) crystal structure with different lattice parameters. For CdS, a = b = 0.413 nm, c = 0.671 nm. For $MoS_2$, a = b = 0.316 nm, c = 1.229 nm. Consequently, for growth of $(00\bar{2})$ $MoS_2$ on (002) CdS, the lattice mismatch between CdS and $MoS_2$ would be as high as 23%. On the other hand, no dislocations were observed at the interface. The absence of misfit dislocations is attributed to the difficulty in nucleation of dislocations in nanostructures.[29, 30]

The hysteresis loops of CdS NWs and $MoS_2$/CdS NWs have been obtained by a superconducting quantum interference device (SQUID) (Figure 2a-c and supplementary material Fig. S2) with different cation exchange time (3 s, 6 s, 9 s, 12 s, 1 min, 5 min, 10 min, 30 min, 60 min, 2 hr, and 6 hr). After 6 hr cation exchange, CdS NWs ($M_s = 1.135 \times 10^{-3}$ (emu/g), $H_{c\perp}$ = 215.8 Oe) transform completely to $MoS_2$ NWs ($M_s = 5.53 \times 10^{-3}$ (emu/g), $H_{c\perp}$ = 149.9 Oe) and both of them are ferromagnetic. Obviously, 1 min growth of $MoS_2$ on CdS NWs leads to the strongest saturation magnetization intensity. In addition, the low temperature (T = 4 K) hysteresis loop shows a higher coercivity ($H_{c\perp}$ = 973Oe) and saturation magnetization intensity ($M_s = 4.43 \times 10^{-2}$ emu/g) (Figure S3) than the coercivity ($H_{c\perp}$ = 200Oe) and saturation magnetization intensity ($M_s = 1.3 \times 10^{-2}$ emu/g) at room temperature. Figure 2d reveals that the $M_s$ shows a trend to decrease with $MoS_2$ thickness.

For further exploration of the origin and nature of the observed magnetization (M), the knowledge of the electronic structure and atom-resolved magnetization in the MoS$_2$/CdS heterostructure is essential. Therefore, we have performed self-consistent spin-polarized electronic structure calculations within the DFT-GGA.[23] The atomic model shown in Figure 1d, which generates the simulated STEM-ADF image (Figure 1c) that fits best to the real STEM ADF image (Figure 1b), has been used in the DFT-GGA calculation.

A previous DFT-GGA calculation shows that free standing few-layer thin films of MoS$_2$ have no magnetic moment.[31] In contrast, the present calculation shows that the MoS$_2$ monolayer next to the CdS substrate becomes ferromagnetic with a Mo magnetic moment ($m_s$) of ~0.5 µB and a large electronic state spin polarization (P) of ~63 % at the Fermi level (Figure 3d). Spin polarization P = $(N_\uparrow - N_\downarrow)/(N_\uparrow + N_\downarrow)$ where $N_\uparrow$ and $N_\downarrow$ are the spin-up and spin-down densities of states (DOSs) at the Fermi level (EF), respectively. The Mo magnetic moment and spin polarization are greatly reduced in the second (ms =0.01µB, P = 15 %) and third ($m_s$ = 0.00 µB, P = 2 %) monolayers of MoS$_2$ away from the interface, as demonstrated by the calculated Mo-decomposed spin-resolved densities of states (DOSs) displayed in Figure 3d (and also Figure S5 in Supplementary Material). Experimentally, the thinner MoS$_2$ layers are observed to possess a higher magnetization ($M_s$), and this is consistent with the results of the DFT calculation. Clearly, both experimental data and DFT calculation indicate that the interfacial MoS$_2$ layer has the predominant contribution to the MoS$_2$/CdS magnetic properties.

The observed dependence of magnetization on the thickness of the MoS$_2$ layer on the CdS

NW is consistent with the theoretical finding of the magnetism in the MoS$_2$ monolayer at the MoS$_2$/CdS interface as revealed by our atomic model (Figures 1d and 3a). Raman peaks ($E_{2g}^1$ and A$_{1g}$) of the MoS$_2$ layer (Supplementary Material Figure S1) are clearly seen after 2 min of cation exchange, indicating that the quality of the MoS$_2$ structure improves as the thickness increases. Indeed, our Raman results show that our MoS$_2$ layer has a fine 2-D structure on the CdS NW with clear in-plane and out-of-plane vibration modes. This explains that the few-layer thick MoS$_2$ structures have poor magnetic property. For example, for very few-layer thick MoS$_2$ films on CdS NWs [3s (0.5 layer), 6s (one layer), 9s (1.5 layer), and 12 s (2 layers)], the M$_s$ values are lower than that for 5 nm (5min for reaction) thick MoS$_2$ on CdS NWs. For MoS$_2$ with an improved structure, the interfacial magnetism starts to dominate the magnetic properties. On the other hand, for MoS$_2$ shells of thicker than 5 nm on CdS NWs, the influence of the interfacial magnetism becomes reduced. As a result, the magnetization decreases with increasing MoS$_2$ thickness. Overall, as the CdS NW gradually transforms to the MoS$_2$ NW, the maximum magnetization moment (M$_{max}$ or M$_S$) increases first sharply with the formation of thin shell of MoS$_2$, then decreases rapidly. This phenomenon is similar to the variation of the photoluminescence intensity of MoS$_2$ with thickness.[32] Furthermore, the magnetic force microscopy (MFM) data also show that the magnetic property of MoS$_2$ depends on the number of layers.[33]

The calculated spin density distribution in the MoS$_2$/CdS heterostructure is exhibited in Figure 3a-b, and that of the Mo atom in the interfacial MoS$_2$ monolayer is displayed in Figure 3c. Clearly, there is significant spin density only in the interfacial MoS$_2$ monolayer (Figure 3a-b), where the spin density distribution of Mo is similar to the $d_{x^2-y^2}$ orbital (Figure 3c). The results indicate that the major contribution comes from Mo d orbitals. This is because the Fermi level is raised above the bottom of the conduction Mo $d_{x^2-y^2}$ and $d_{xy}$ bands in the interfacial monolayer, as revealed by calculated Mo-decomposed spin-resolved DOSs shown in Figure 3d (see also Supplementary Material Figure S5). Interestingly, the calculated spin-resolved DOS curves show that the local DOS of the Cd atoms on the interfacial Cd monolayer is also spin-polarized (Supplementary Material Figure S6), due to the magnetic proximity effect on the Cd atoms by the magnetic Mo atoms via the S atoms across the interface. For example, the spin polarizations for the S atoms on the interface S monolayer and the Cd atoms on the Cd monolayer next to the interface are, respectively, 20 % ($S_{M1}$) and 16 % ($Cd_1$) (see Figure 4b and also Figures S6 and S7 in Supplementary Material).

Figure 4a shows the calculated charge density distribution in the MoS$_2$/CdS heterostructure as modelled by the atomic model shown in Fig. 3b. For the sake of discussion, the different atomic locations in the vicinity of the MoS$_2$/CdS interface are labelled, as shown in Figure 4b. The charge density distribution in the region from the $S_{M4}$ to $Cd_2$ atom layer is rather flat, having a similar isosurface of isovalue = 0.139 e/Å$^3$. Inside the region, the electron density in some areas is higher than the isovalue, indicating some bonding between the S and Cd atoms across the MoS$_2$/CdS interface. This covalent bonding could be attributed to the similar electronegativities of the Mo (1.3) and Cd (1.5) atoms. Remarkably, the calculated DOS curves (Supplementary Material Figure S5) show that this covalent bonding causes some charge transfer from the Mo d bands in the outer $MoS_2$ monolayer ($Mo_3$) to the Mo d bands in the interfacial $MoS_2$ monolayer ($Mo_1$), leading to the partial occupation of the Mo $d_{x^2-y^2}$ and dxy dominant bands of the $Mo_1$ atoms. This charge transfer is also seen from the calculated numbers of the valence electrons on the $Mo_1$ (6.38), $Mo_2$ (6.17) and $Mo_3$ (6.02) atoms (Supplementary Material Table 1). As a result, the strong exchange interaction among the Mo d electrons then gives rise to the formation of the $Mo_1$ spin magnetic moment, and hence the ferromagnetic $MoS_2$ monolayer next to the $MoS_2$/CdS.

**CONCLUSIONS**

In summary, magnetic $MoS_2$ thin films have been grown on CdS NWs by cation exchange at room temperature. The magnetism in the $MoS_2$/CdS heterostructure was shown to stem from the ferromagnetic $MoS_2$ monolayer next to the $MoS_2$/CdS interface by first-principles DFT-GGA calculations. The ferromagnetism was attributed to the partial occupation of the Mo $d_{x^2-y^2}/d_{xy}$ conduction band in the interfacial $MoS_2$ monolayer caused by the mixed covalent-ionic bonding among the $MoS_2$ and CdS monolayers near the $MoS_2$/CdS interface. This work thus offers a new paradigm for fabrication of the transition metal dichalcogenide-based magnetic semiconductor multilayers for applications in spintronic devices such as tunnel junctions with high magnetoresistance.

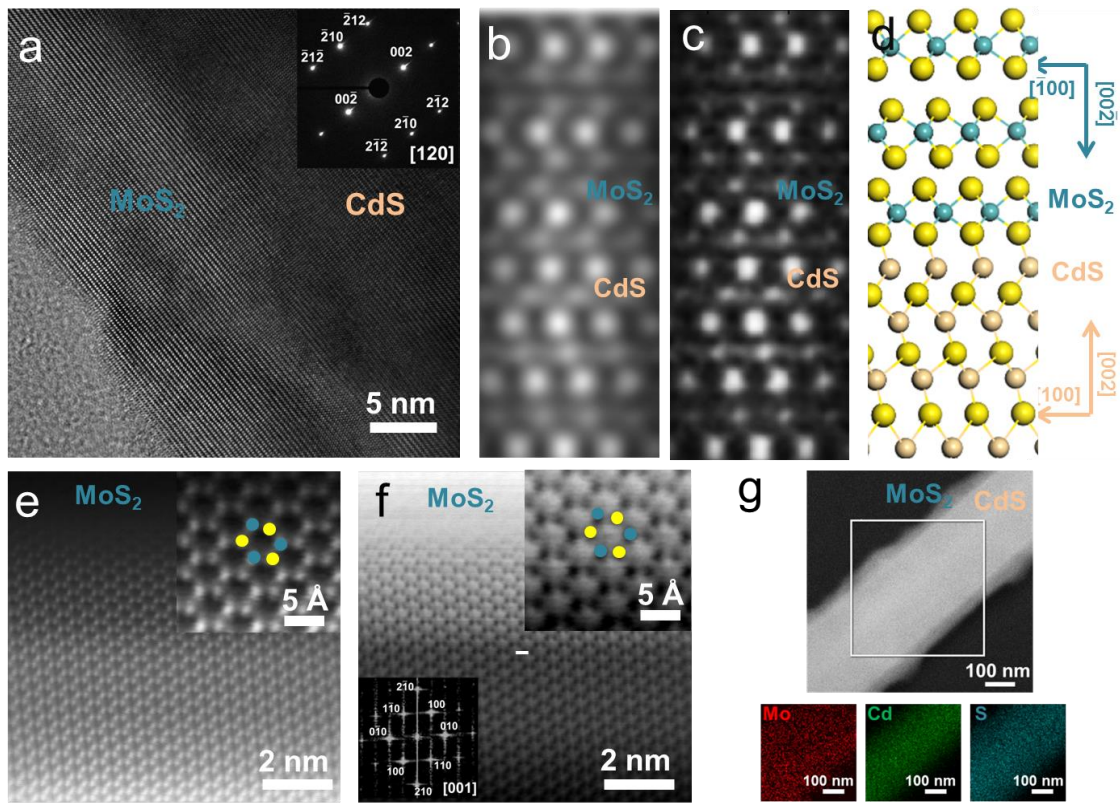

Figure 1. Heterojunction area analysis of MoS$_2$/CdS NWs after cation exchange. (a) The high resolution TEM image of heterostructure of MoS$_2$/CdS at junction area (after 3 min reaction). Inset is the corresponding FFT pattern. (b) The STEM-ADF image of the MoS$_2$/CdS heterostructure (after 3 min reaction). (c) The simulated STEM-ADF image of the MoS$_2$/CdS heterostructure. (d) Atomic model of the MoS$_2$/CdS junction. (e) STEM-ADF image of MoS$_2$ region (after 3 min reaction). Upper right inset is the magnified image of e. (f) STEM-ABF image of MoS$_2$ region (after 3 min reaction). Upper right and lower left insets are the magnified image and FFT pattern, respectively. The blue and yellow dots represent Mo and S atoms, respectively. (g) STEM image and EDS mapping of CdS NW after 10 min Mo ion exchange.

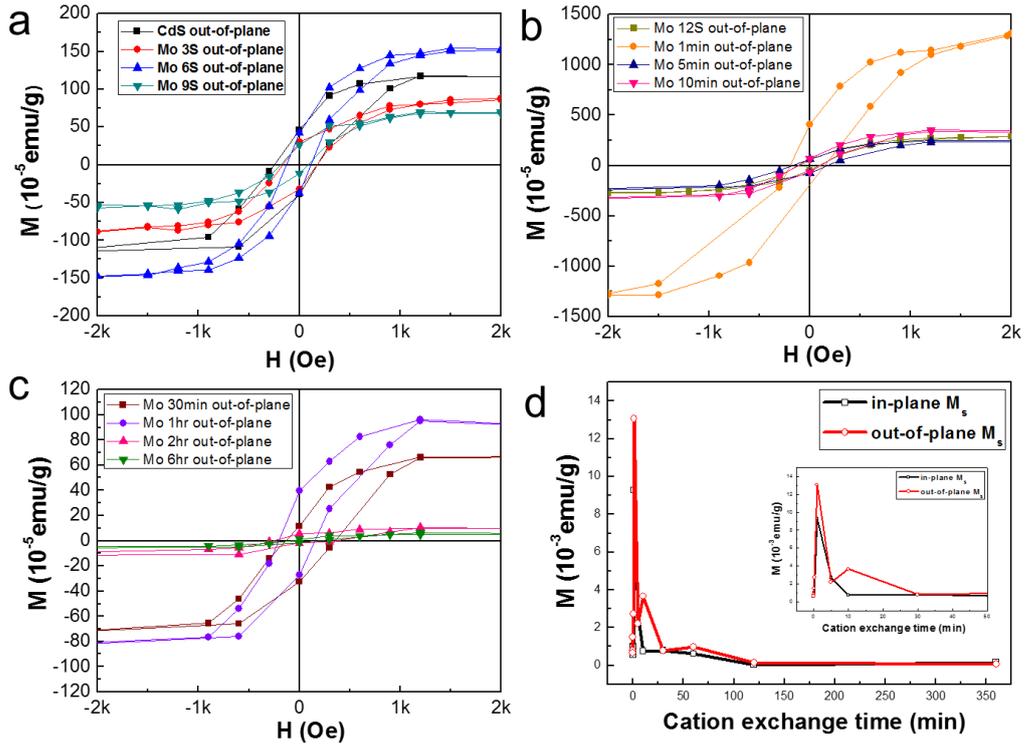

Figure 2. Magnetic measurements of MoS$_2$/CdS NWs. (a)-(c) Hysteresis loops of CdS NWs and MoS$_2$/CdS NWs in out of plane direction. M is magnetization and H is applied field. (d) Saturation magnetization of MoS$_2$/CdS NWs as a function of cation exchange time.

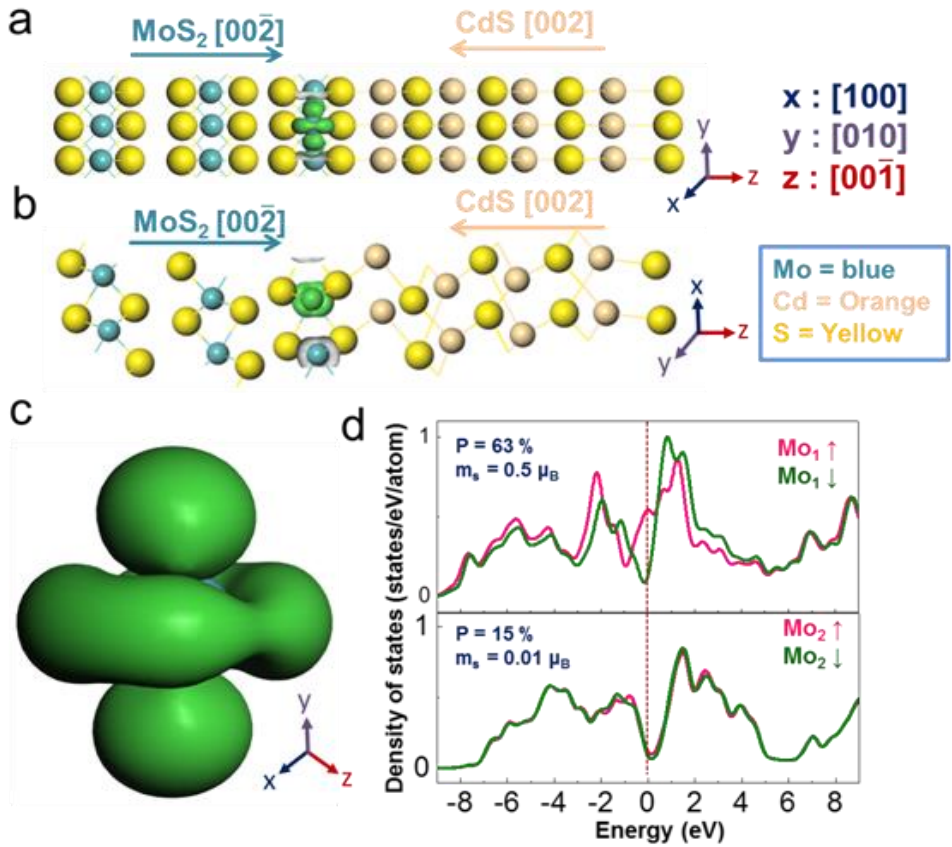

Figure 3. Atomic structure and DFT calculation of magnetic property of the MoS$_2$/CdS heterotructure. Atomic models of the MoS$_2$/CdS heterostructure with (00$\bar{2}$) MoS$_2$ connecting with (002) CdS along (a) [010] and (b) [110] zones axes. The calculated spin density is concentrated at the first layer of Mo atoms from the junction. (c) The shape (iso-surface) of spin density of the Mo atom on the interfacial MoS$_2$ monolayer. (d) The spin-resolved DOSs of the Mo atoms on the first and second layers to the interface (upper and lower panels). Red dash line indicates the Fermi level.

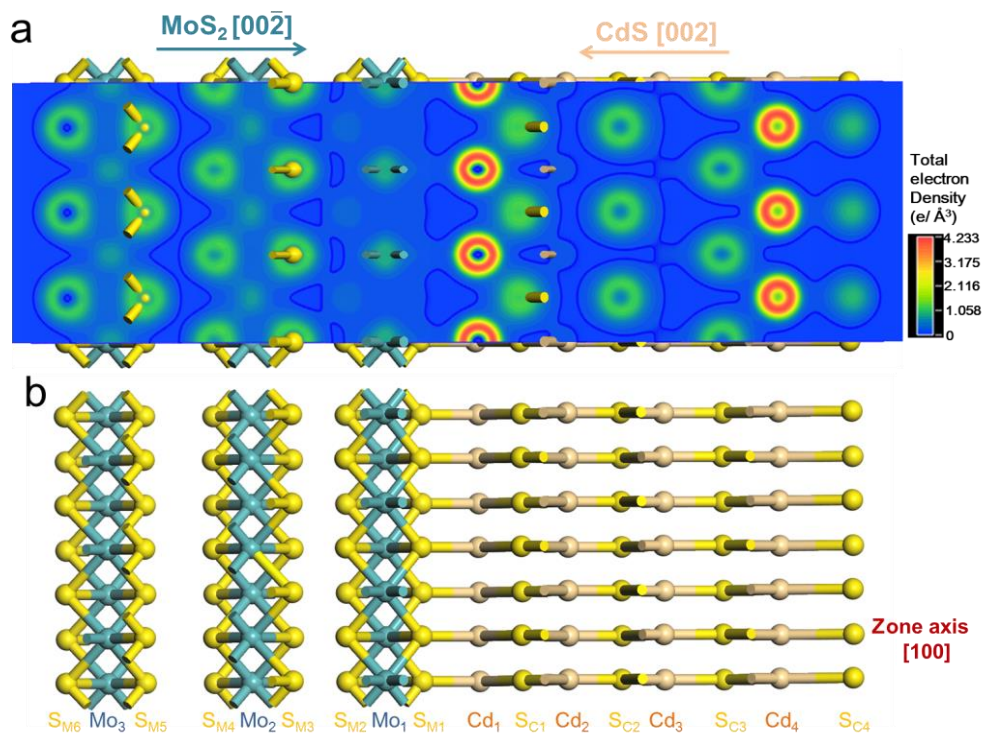

Figure 4. Electron density distribution of $MoS_2$/CdS heterostructure. (a) Total electron density distribution at $MoS_2$/CdS heterostructure and shown with isosurface (blue line, isovalue = 0.13876 (e/Å$^3$)). (b) Atomic model of the $MoS_2$/CdS heterostructure with [100] zone axis.

**ASSOCIATED CONTENT**

**Supporting Information**.

This material contains additional figures displaying measured Raman spectra, SEM images and hysteresis loops as well as calculated atom- and orbital-decomposed densities of states; additional tables listing calculated occupation numbers of atomic orbitals, and also structural parameters of the atomic model adopted in the SEM image simulations and DFT-GGA calculations (pdf).

# AUTHOR INFORMATION

**Corresponding Author**

Correspondence and requests for materials should be addressed to G. Y. Guo and L. J. Chen. (gyguo@phys.ntu.edu.tw, ljchen@mx.nthu.edu.tw)

# ACKNOWLEDGMENT

We thank K. L. Lin at National Nano Device Laboratories for discussion of experimental results. The authors are grateful for the research support from the Ministry of Science and Technology of R.O.C. under project numbers MOST 103-2221-E-007-003 and MOST 104-2112-M-002-002-MY3, Academia Sinica Thematic Research Program and National Center of Theoretical Sciences.

# Supplementary Materials for

**Magnetic MoS$_2$ Interface Monolayer on CdS Nanowire by Cation Exchange**

Chih-Shan Tan[1], Yu-Jung Lu[2], Chun-Chi Chen[3], Pei-Hsuan Liu[1], Shangjr Gwo[2], Guang-Yu Guo[4,5,*], Lih-Juann Chen[1,*]

[1]Department of Materials Science and Engineering, National Tsing Hua University, Hsinchu, Taiwan, 30043, R.O.C.

[2]Department of Physics, National Tsing Hua University, Hsinchu, Taiwan, 30043, R.O.C.

[3]National Nano Device Laboratories, National Applied Research Laboratories, 26, Prosperity Road I, Hsinchu, Taiwan, 30078, R.O.C.

[4]Department of Physics, National Taiwan University, Taipei, Taiwan, 10617, R.O.C

[5]Physics Division, National Center for Theoretical Sciences, Hsinchu, Taiwan, 30078, R.O.C

correspondence to: gyguo@phys.ntu.edu.tw, ljchen@mx.nthu.edu.tw

**This PDF file includes:**

Figures S1 to S7

Tables S1 to S3

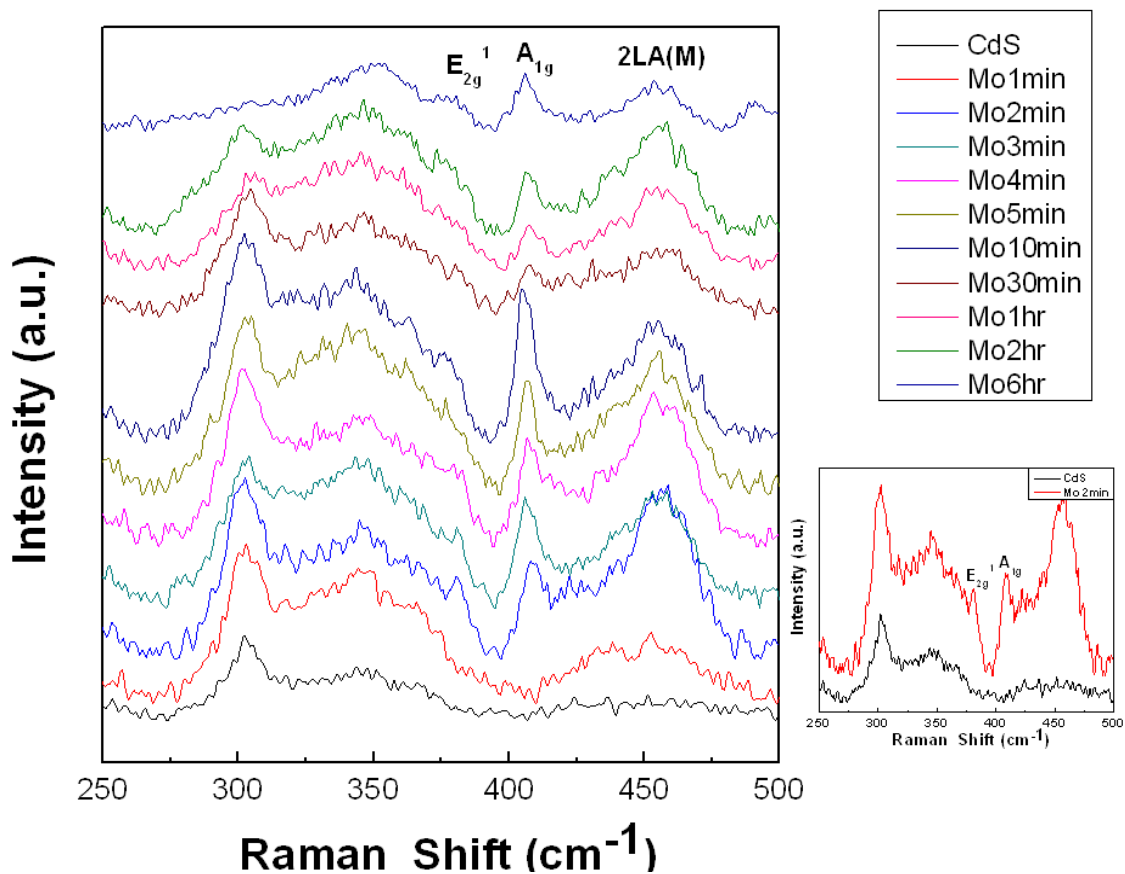

**Figure S1.** Raman spectra following cation exchange from CdS NWs to $MoS_2$ NWs. CdS NWs were dipped in 0.1 M molybdenum ion solution at room temperature with different reaction time.

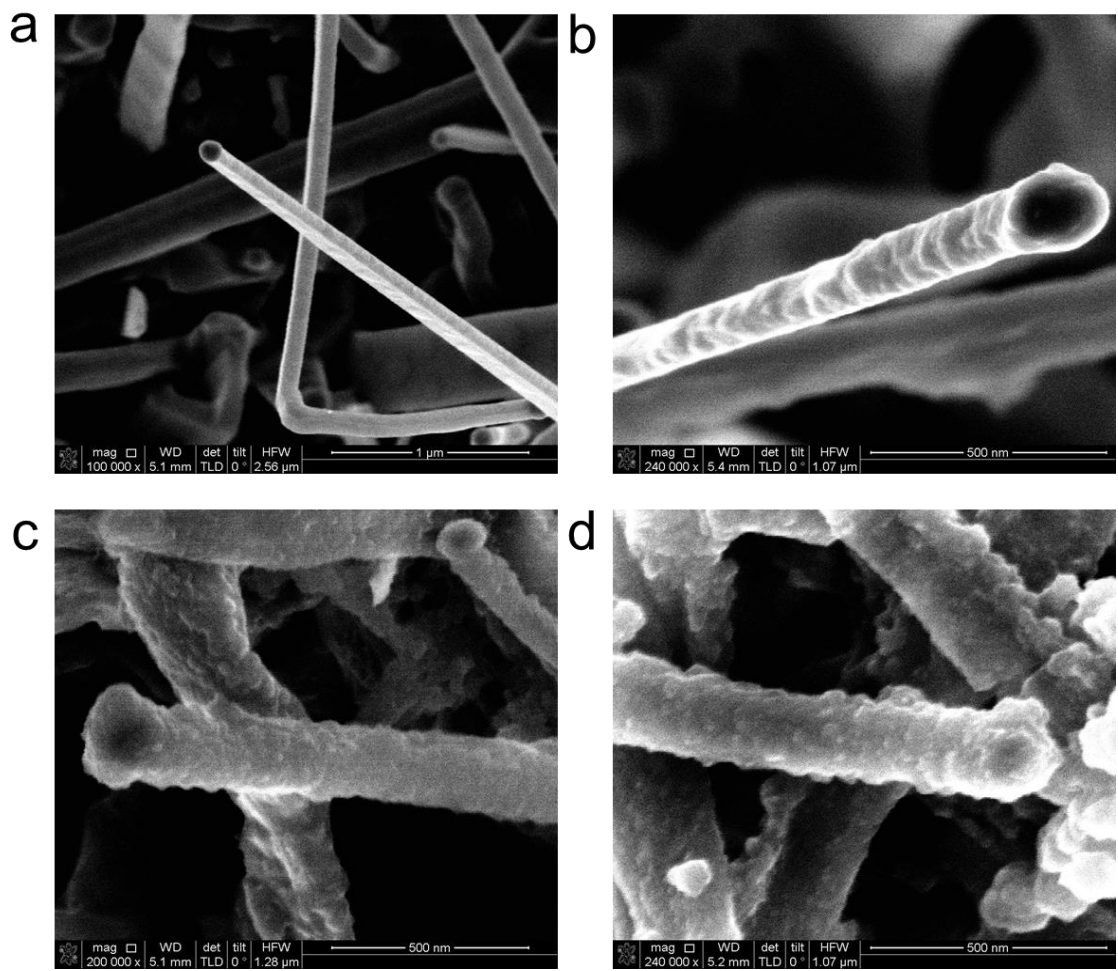

**Figure S2.** Scanning electron microscopy (SEM) images of cation exchange from CdS NWs to MoS$_2$ NWs. (a) CdS NW. (b) CdS NW in Mo ion solution for 10 min. (c) CdS NW in Mo ion solution for 2 hr. (d) CdS NW in Mo ion solution for 6 hr.

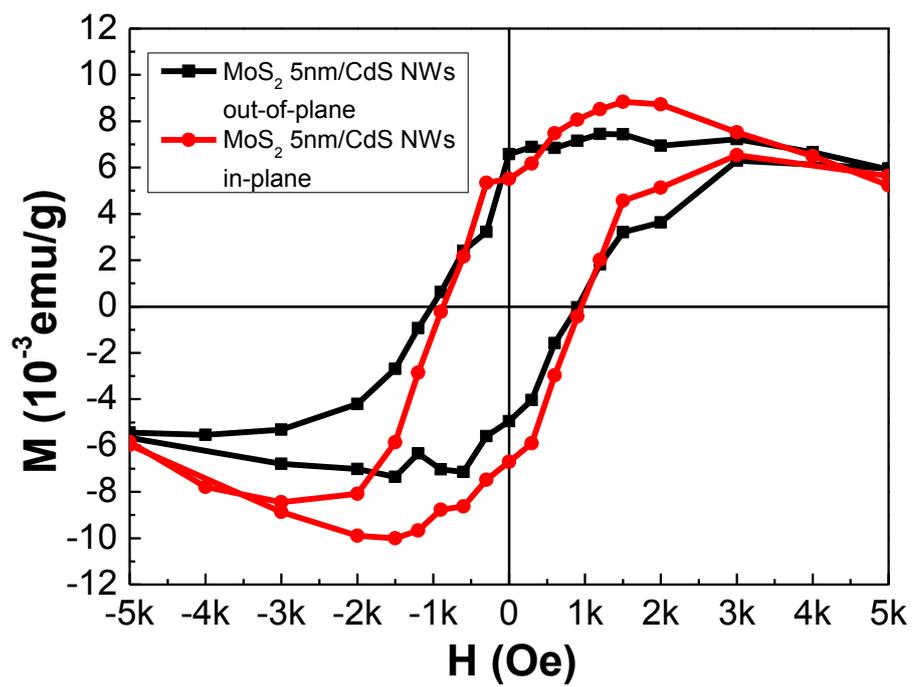

**Figure S3.** The hysteresis loops of MoS$_2$ (5nm)/CdS NWs at 4K.

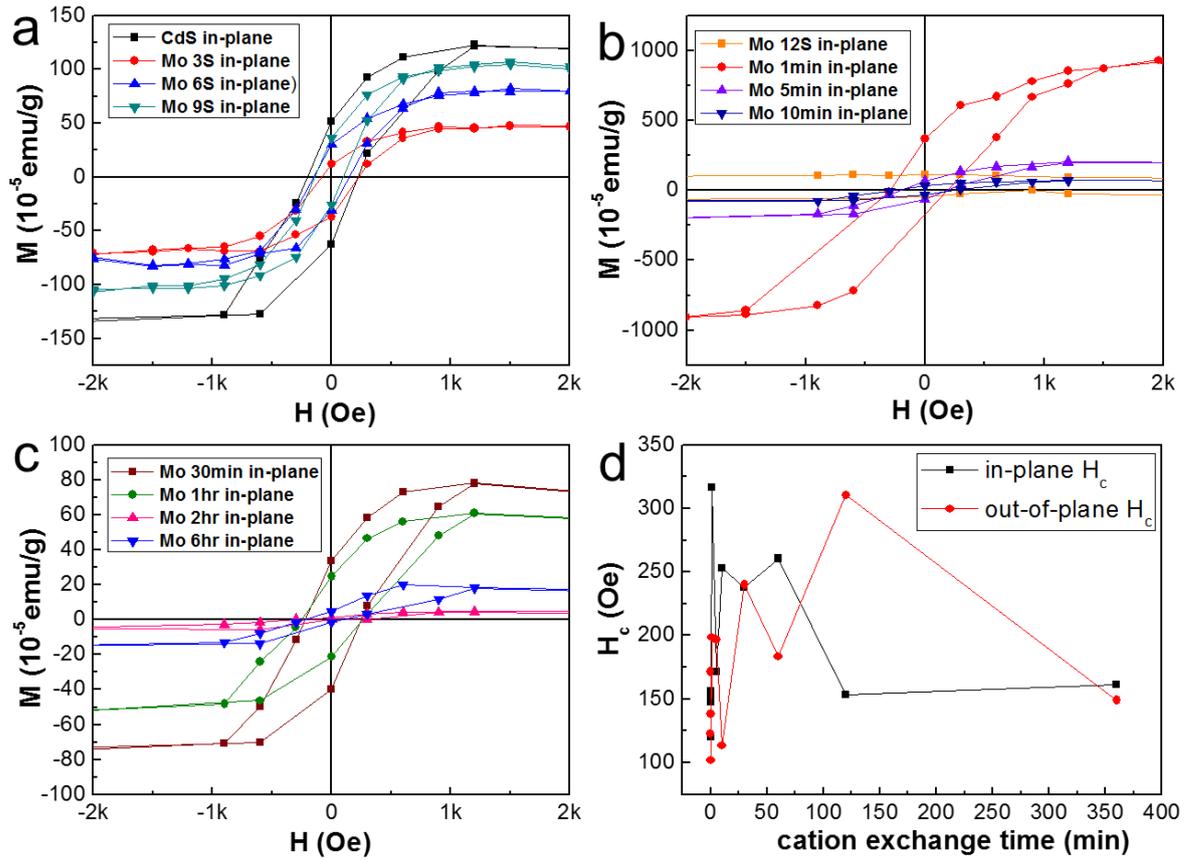

**Figure S4.** The hysteresis loops of CdS NWs and MoS$_2$/CdS NWs at in-plane direction. (a)-(c) CdS NWs after cation exchange for 3 s to 6 hr. (d) The coercivity of CdS NWs transforming to MoS$_2$ NWs with different cation exchange time in in-plane and out-of-plane directions.

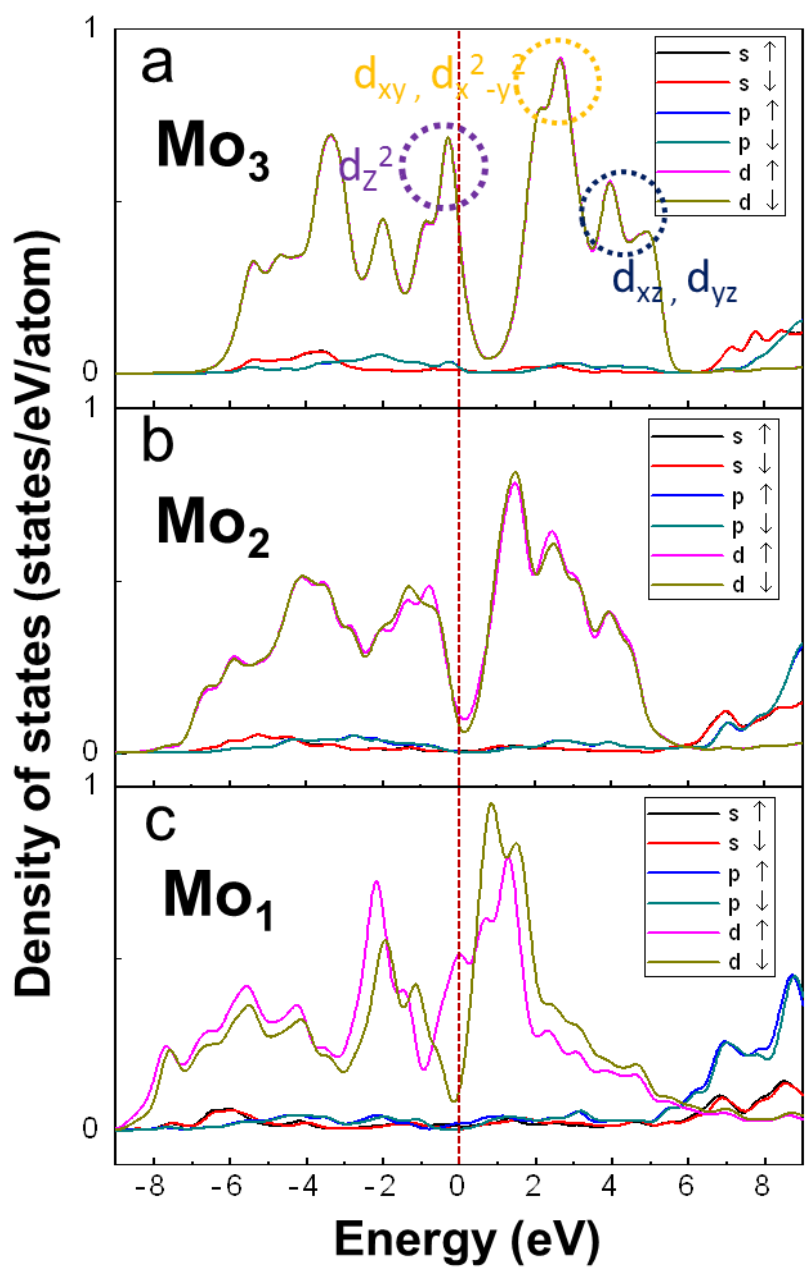

**Figure S5.** Calculated orbital-decomposed spin-up (↑) and spin-down (↓) densities of states (ODOSs) of the Mo atoms at different locations (see Figure 4b for the labelling of the locations) in the $MoS_2$/CdS heterostructure. Red dashed line indicates the Fermi level. The labelling of the atomic positions is shown in Figure 4b.

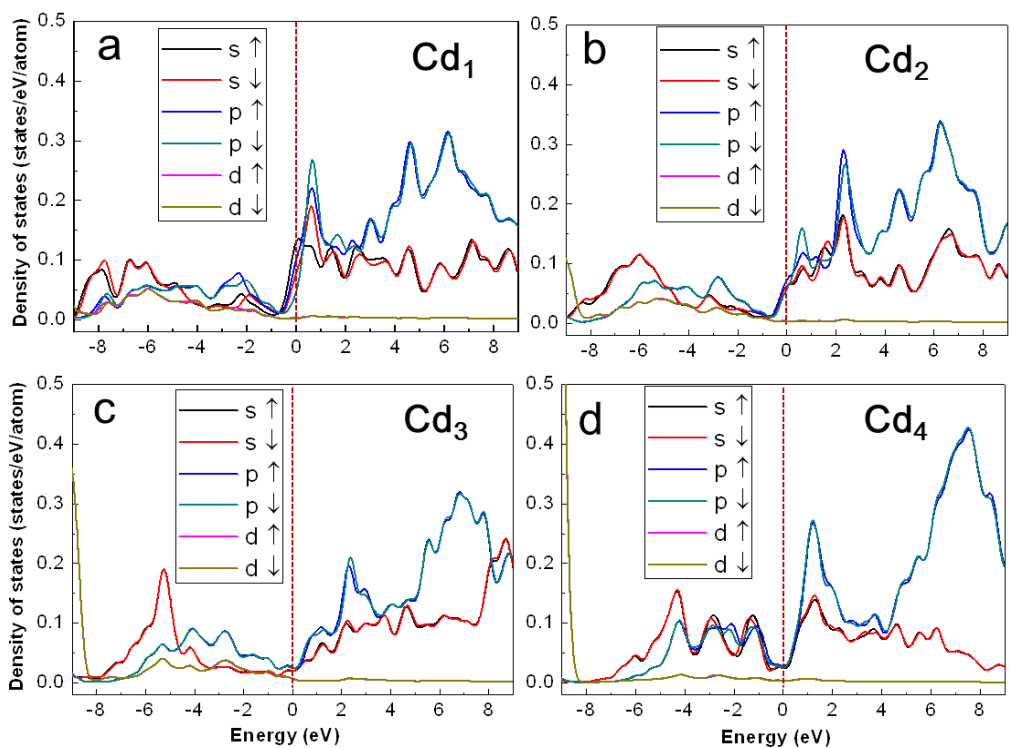

**Figure S6.** Calculated orbital-decomposed spin-up (↑) and spin-down (↓) densities of states (ODOSs) of the Cd atoms at different locations (see Figure 4b for the labelling of the locations) in the $MoS_2$/CdS heterostructure. Red dashed line indicates the Fermi level. The labelling of the atomic positions is shown in Figure 4b.

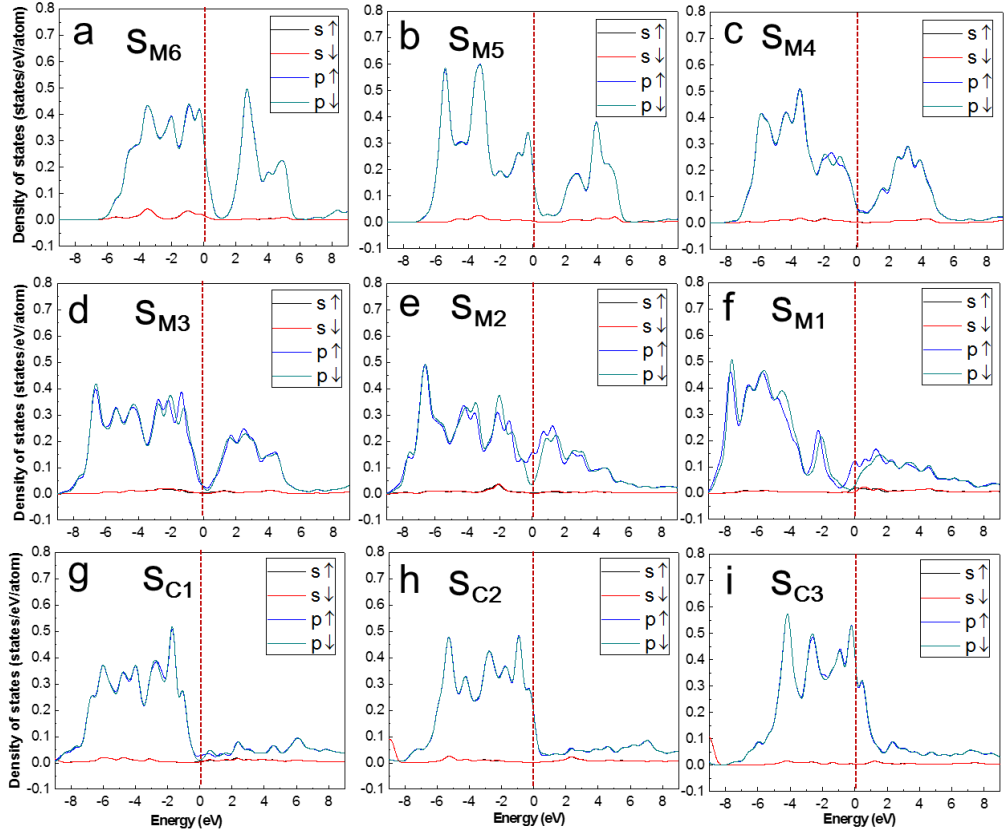

**Figure S7.** Calculated orbital-decomposed spin-up (↑) and spin-down (↓) densities of states (ODOSs) of the S atoms at different locations (see Figure 4b for the labelling of the locations) in the $MoS_2$/CdS heterostructure. Red dashed line at zero energy indicates the Fermi level. The labelling of the atomic positions is shown in Figure 4b.

**Table S1.** The calculated numbers of the valence electrons of the Mo, Cd, and S atoms near the MoS$_2$/CdS interface. The labelling of the atomic positions is shown in Figure 4b.

|     | s electrons | p electrons | d electrons | Total electrons |
| --- | --- | --- | --- | --- |
| Mo1 | 0.63 | 0.42 | 5.33 | 6.38 |
| Mo2 | 0.57 | 0.41 | 5.19 | 6.17 |
| Mo3 | 0.55 | 0.35 | 5.12 | 6.02 |
| Cd1 | 1.13 | 1.02 | 9.96 | 12.11 |
| Cd2 | 1.13 | 0.98 | 9.96 | 12.07 |
| Cd3 | 1.08 | 1 | 9.95 | 12.04 |
| Cd4 | 1.09 | 0.93 | 9.98 | 12 |
| SM6 | 1.95 | 3.66 | 0 | 5.61 |
| SM5 | 1.86 | 4.1 | 0 | 5.96 |
| SM4 | 1.87 | 4.07 | 0 | 5.94 |
| SM3 | 1.87 | 4.14 | 0 | 6.01 |
| SM2 | 1.87 | 4.08 | 0 | 5.95 |
| SM1 | 1.8 | 4.27 | 0 | 6.06 |
| Sc1 | 1.79 | 4.73 | 0 | 6.52 |
| Sc2 | 1.79 | 4.68 | 0 | 6.47 |
| Sc3 | 1.86 | 4.4 | 0 | 6.26 |

**Table S2.** The atomic structure of the supercell model of the MoS$_2$/CdS heterostructure constructed (see the main text) to generate the simulated STEM-ADF image [see Fig. 1 (c)] as shown in Fig. 4. The lattice constants are $a$ = 5.508514 Å, $b$ = 9.908285 Å and $c$ = 47.830078 Å. The angles between the crystalline axes are α = 90°, β = 94.693° and γ = 90°. The unit cell vectors (in the unit of Å) are (5.490045, 0, -0.450712), (0, 9.908285) and (0, 0, 47.830078).

| number | element | Fractional coordinates (u, v, w) | | |
|---|---|---|---|---|
| 1 | S | 0.031054 | 0.000000 | 0.365889 |
| 2 | S | 0.744559 | 0.000000 | 0.465031 |
| 3 | S | 0.696706 | 0.000000 | 0.407148 |
| 4 | S | 0.937715 | 0.000000 | 0.305079 |
| 5 | S | 0.569778 | 0.000000 | 0.252281 |
| 6 | S | 0.480304 | 0.000000 | 0.191234 |
| 7 | S | 0.595002 | 0.000000 | 0.542408 |
| 8 | S | 0.450019 | 0.000000 | 0.618776 |
| 9 | S | 0.955831 | 0.000000 | 0.697003 |
| 10 | S | 0.377858 | 0.000000 | 0.796940 |
| 11 | S | 0.531054 | 0.166667 | 0.365889 |
| 12 | S | 0.244559 | 0.166667 | 0.465031 |
| 13 | S | 0.196706 | 0.166667 | 0.407148 |
| 14 | S | 0.437715 | 0.166667 | 0.305079 |
| 15 | S | 0.069778 | 0.166667 | 0.252281 |
| 16 | S | 0.980304 | 0.166667 | 0.191234 |
| 17 | S | 0.095002 | 0.166667 | 0.542408 |
| 18 | S | 0.950019 | 0.166667 | 0.618776 |
| 19 | S | 0.455831 | 0.166667 | 0.697003 |
| 20 | S | 0.877858 | 0.166667 | 0.796940 |
| 21 | S | 0.031054 | 0.333333 | 0.365889 |
| 22 | S | 0.744559 | 0.333333 | 0.465031 |
| 23 | S | 0.696706 | 0.333333 | 0.407148 |
| 24 | S | 0.937715 | 0.333333 | 0.305079 |
| 25 | S | 0.569778 | 0.333333 | 0.252281 |
| 26 | S | 0.480304 | 0.333333 | 0.191234 |
| 27 | S | 0.595002 | 0.333333 | 0.542408 |
| 28 | S | 0.450019 | 0.333333 | 0.618776 |
| 29 | S | 0.955831 | 0.333333 | 0.697003 |
| 30 | S | 0.377858 | 0.333333 | 0.796940 |
| 31 | S | 0.531054 | 0.500000 | 0.365889 |
| 32 | S | 0.244559 | 0.500000 | 0.465031 |
| 33 | S | 0.196706 | 0.500000 | 0.407148 |
| 34 | S | 0.437715 | 0.500000 | 0.305079 |
| 35 | S | 0.069778 | 0.500000 | 0.252281 |
| 36 | S | 0.980304 | 0.500000 | 0.191234 |
| 37 | S | 0.095002 | 0.500000 | 0.542408 |
| 38 | S | 0.950019 | 0.500000 | 0.618776 |
| 39 | S | 0.455831 | 0.500000 | 0.697003 |
| 40 | S | 0.877858 | 0.500000 | 0.796940 |
| 41 | S | 0.031054 | 0.666667 | 0.365889 |
| 42 | S | 0.744559 | 0.666667 | 0.465031 |
| 43 | S | 0.696706 | 0.666667 | 0.407148 |
| 44 | S | 0.937715 | 0.666667 | 0.305079 |
| 45 | S | 0.569778 | 0.666667 | 0.252281 |
| 46 | S | 0.480304 | 0.666667 | 0.191234 |
| 47 | S | 0.595002 | 0.666667 | 0.542408 |
| 48 | S | 0.450019 | 0.666667 | 0.618776 |
| 49 | S | 0.955831 | 0.666667 | 0.697003 |
| 50 | S | 0.377858 | 0.666667 | 0.796940 |
| 51 | S | 0.531054 | 0.833333 | 0.365889 |
| 52 | S | 0.244559 | 0.833333 | 0.465031 |
| 53 | S | 0.196706 | 0.833333 | 0.407148 |
| 54 | S | 0.437715 | 0.833333 | 0.305079 |
| 55 | S | 0.069778 | 0.833333 | 0.252281 |
| 56 | S | 0.980304 | 0.833333 | 0.191234 |
| 57 | S | 0.095002 | 0.833333 | 0.542408 |
| 58 | S | 0.950019 | 0.833333 | 0.618776 |
| 59 | S | 0.455831 | 0.833333 | 0.697003 |
| 60 | S | 0.877858 | 0.833333 | 0.796940 |
| 61 | Mo | 0.116231 | 0.000000 | 0.437692 |
| 62 | Mo | 0.641404 | 0.000000 | 0.334100 |
| 63 | Mo | 0.867034 | 0.000000 | 0.223515 |
| 64 | Mo | 0.616231 | 0.166667 | 0.437692 |
| 65 | Mo | 0.141404 | 0.166667 | 0.334100 |
| 66 | Mo | 0.367034 | 0.166667 | 0.223515 |
| 67 | Mo | 0.116231 | 0.333333 | 0.437692 |
| 68 | Mo | 0.641404 | 0.333333 | 0.334100 |
| 69 | Mo | 0.867034 | 0.333333 | 0.223515 |
| 70 | Mo | 0.616231 | 0.500000 | 0.437692 |
| 71 | Mo | 0.141404 | 0.500000 | 0.334100 |
| 72 | Mo | 0.367034 | 0.500000 | 0.223515 |
| 73 | Mo | 0.116231 | 0.666667 | 0.437692 |
| 74 | Mo | 0.641404 | 0.666667 | 0.334100 |
| 75 | Mo | 0.867034 | 0.666667 | 0.223515 |
| 76 | Mo | 0.616231 | 0.833333 | 0.437692 |
| 77 | Mo | 0.141404 | 0.833333 | 0.334100 |
| 78 | Mo | 0.367034 | 0.833333 | 0.223515 |
| 79 | Cd | 0.998515 | 0.000000 | 0.509028 |
| 80 | Cd | 0.221681 | 0.000000 | 0.577774 |
| 81 | Cd | 0.766340 | 0.000000 | 0.651496 |
| 82 | Cd | 0.450554 | 0.000000 | 0.740489 |
| 83 | Cd | 0.498515 | 0.166667 | 0.509028 |
| 84 | Cd | 0.721681 | 0.166667 | 0.577774 |
| 85 | Cd | 0.266340 | 0.166667 | 0.651496 |
| 86 | Cd | 0.950554 | 0.166667 | 0.740489 |
| 87 | Cd | 0.998515 | 0.333333 | 0.509028 |
| 88 | Cd | 0.221681 | 0.333333 | 0.577774 |
| 89 | Cd | 0.766340 | 0.333333 | 0.651496 |
| 90 | Cd | 0.450554 | 0.333333 | 0.740489 |
| 91 | Cd | 0.498515 | 0.500000 | 0.509028 |
| 92 | Cd | 0.721681 | 0.500000 | 0.577774 |
| 93 | Cd | 0.266340 | 0.500000 | 0.651496 |
| 94 | Cd | 0.950554 | 0.500000 | 0.740489 |
| 95 | Cd | 0.998515 | 0.666667 | 0.509028 |
| 96 | Cd | 0.221681 | 0.666667 | 0.577774 |
| 97 | Cd | 0.766340 | 0.666667 | 0.651496 |
| 98 | Cd | 0.450554 | 0.666667 | 0.740489 |
| 99 | Cd | 0.498515 | 0.833333 | 0.509028 |
| 100 | Cd | 0.721681 | 0.833333 | 0.577774 |
| 101 | Cd | 0.266340 | 0.833333 | 0.651496 |
| 102 | Cd | 0.950554 | 0.833333 | 0.740489 |

**Table S3.** The atomic structure of the supercell model of the MoS$_2$/CdS heterostructure constructed for the electronic structure calculations (see the main text) as shown in Fig. 3 as well as Fig. S5-S7. The lattice constants are $a$ = 5.508514 Å, $b$ = 3.302762 Å and $c$ = 47.830078 Å. The angles between the crystalline axes are α = 90°, β = 94.693° and γ = 90°. The unit cell vectors (in the unit of Å) are (5.490045, 0, -0.450712), (0, 3.302762) and (0, 0, 47.830078). The structure is the same as Table S2 except that lattice constant $b$ is three times smaller.

| number | element | Fractional coordinates (u, v, w) | | |
|---|---|---|---|---|
| 1 | S | 0.031054 | 0.000000 | -0.634111 |
| 2 | S | -0.255441 | 0.000000 | -0.534969 |
| 3 | S | -0.303294 | 0.000000 | -0.592852 |
| 4 | S | -0.062285 | 0.000000 | -0.694921 |
| 5 | S | -0.430222 | 0.000000 | -0.747719 |
| 6 | S | -0.519696 | 0.000000 | -0.808766 |
| 7 | S | -0.404998 | 0.000000 | -0.457592 |
| 8 | S | -0.549981 | 0.000000 | -0.381224 |
| 9 | S | -0.044169 | 0.000000 | -0.302997 |
| 10 | S | -0.622142 | 0.000000 | -0.203060 |
| 11 | S | 0.531054 | 0.500000 | -0.634111 |
| 12 | S | 0.244559 | 0.500000 | -0.534969 |
| 13 | S | 0.196706 | 0.500000 | -0.592852 |
| 14 | S | 0.437715 | 0.500000 | -0.694921 |
| 15 | S | 0.069778 | 0.500000 | -0.747719 |
| 16 | S | -0.019696 | 0.500000 | -0.808766 |
| 17 | S | 0.095002 | 0.500000 | -0.457592 |
| 18 | S | -0.049981 | 0.500000 | -0.381224 |
| 19 | S | 0.455831 | 0.500000 | -0.302997 |
| 20 | S | -0.122142 | 0.500000 | -0.203060 |
| 21 | Mo | -0.883769 | 0.000000 | -0.562308 |
| 22 | Mo | -0.358596 | 0.000000 | -0.665900 |

| 23 | Mo | -0.132966 | 0.000000 | -0.776485 |
| 24 | Mo | -0.383769 | 0.500000 | -0.562308 |
| 25 | Mo | 0.141404 | 0.500000 | -0.665900 |
| 26 | Mo | 0.367034 | 0.500000 | -0.776485 |
| 27 | Cd | -0.001485 | 0.000000 | -0.490972 |
| 28 | Cd | -0.778319 | 0.000000 | -0.422226 |
| 29 | Cd | -0.233660 | 0.000000 | -0.348504 |
| 30 | Cd | -0.549446 | 0.000000 | -0.259511 |
| 31 | Cd | 0.498515 | 0.500000 | -0.490972 |
| 32 | Cd | -0.278319 | 0.500000 | -0.422226 |
| 33 | Cd | 0.266340 | 0.500000 | -0.348504 |
| 34 | Cd | -0.049446 | 0.500000 | -0.259511 |